\documentstyle[multicol,epsfig,epsf,aps,prl]{revtex}
\begin{document}
\draft
\tightenlines
\title{Multi-Channel Transport in Disordered Medium under Generic Scattering 
Conditions}
\author{Pragya Shukla$^{1,2}$ and Inder P. Batra$^{1}$}
\address{ Department of Physics,
University of Illinois at Chicago, Chicago, Illinois-60607, USA$^{1}$ \\
Department of Physics, Indian Institute of Technology, Kharagpur, 
India$^{2}$}
\onecolumn
\date{\today}
\maketitle
\widetext
\begin{abstract}
% .

        Our study of the evolution of transmission eigenvalues, due to
changes in various physical parameters in a disordered region of arbitrary 
dimensions, results in a generalization of the celebrated DMPK equation. 
The evolution is shown to be governed by a single complexity parameter which 
implies a  deep level of universality of transport phenomena through a wide 
range of disordered regions. We also find that the interaction among 
eigenvalues is of many body type that has important 
consequences for the statistical behavior of transport properties.

\end{abstract}
                                                 
%..
%\begin{multicols}{2}

% .....

\section{Introduction}

	Recent advances in nanotechnology and quantum information theory have 
motivated an extensive research on the topic of electronic transport through 
disordered regions \cite{been,been1,sd}. The random distribution of impurities 
in such systems give rise to fluctuations in transport properties from sample 
to sample. The fluctuations can also be observed in a single disordered sample 
under an external perturbation or a slight variation of a system parameter 
\cite{ww,mw}. As a result, the information about the statistical behavior of 
transport properties is of great importance.

	A variety of transport properties can be formulated in terms of the  
eigenvalues of transmission matrix of the region. The knowledge of the statistical 
behavior of transmission eigenvalues is therefore very useful in the    
statistical analysis of transport properties. 
This motivates us to study the joint probability distribution of transmission 
eigenvalues. Previous attempts in 
this direction have resulted in the well-known DMPK equation which 
describes the statistical evolution of transmission eigenvalues with 
respect to changing length of the medium \cite{been1,mpk}. Various 
assumptions made in its derivation, however, restrict its applicability to 
quasi one dimensional systems or under specific scattering 
conditions \cite{been,been1,mk}. This being the only analytical tool available so 
far to describe the full distribution of transport properties, a 
generalization of DMPK equation for higher 
dimensions and under generic scattering conditions is required. (The other 
analytical method based on nonlinear sigma model provides information 
only about the moments of the transport properties). 
Further the transport properties are also sensitive to changes in other system 
parameters besides length e.g., boundary conditions, disorder strength and 
dimensionality. It would also be  desirable if the generalized equation 
contains explicit information about the effect of various system parameters  
on the distribution. Our study in this paper is aimed at such a generalization.    
    
	The derivation of DMPK equation is based on the transfer matrix approach, 
applied to a conductor placed between two perfect leads of finite width.
A transfer matrix relates the wave amplitudes on the right of the region to the 
left.  The scattering of waves by randomly distributed impurities leads to a
randomization of transferred amplitudes. The transfer matrix 
can therefore be modeled by a random matrix, that is, a matrix with all or 
some of its elements as randomly distributed \cite{been,been1,im,mpd}. 
The idea to use random matrix approach originates from the  
 universality of the conductance fluctuations, observed in the metallic regime of
different disordered systems. The universality suggested the possibility
of formulating a theory which is system-independent; the suggestion motivated 
the use of standard random matrix ensembles \cite{been,been1} 
as models for  the ensembles of transfer matrices in the metallic regime. 
These ensembles are obtained by using maximum entropy hypothesis under a single constraint of 
fixed eigenvalue density; their matrix elements are of almost same strength 
and the statistical behavior, being governed only by underlying 
symmetry, is universal in nature \cite{been,been1}. These ensemble can therefore 
serve as good models for the transfer matrix in a metallic regime or for quasi one 
dimensional conductors where the flux incident on one 
channel can be assumed to be transmitted with the same probability into 
all outgoing channels (known as the isotropic assumption). 
The DMPK equation was also derived by using the standard random matrix model  
for the transfer matrix of a small length of the disordered region. 
As expected, the equation has been very successful in predicting the 
behavior of transport properties in metallic regime or 
for quasi one dimensional conductors.

The standard random matrix ensembles are not appropriate tools to model the 
 transfer matrix of a disordered region under generic scattering conditions. 
This is confirmed by the observed failure of DMPK equation beyond metallic 
regime, in conductors of dimensions higher than one, or under inelastic scattering 
conditions. The generic scattering conditions may cause a preferential as well as 
multiple interactions among  channels which would lead to varying degree 
of correlations among transfer matrix elements. 
The equation governing the evolution of the transmission eigenvalues therefore 
should be based on a model of transfer matrices free of isotropy constraints. 
In this paper, we consider such a model  and study the evolution of 
the transmission eigenvalues due to change of various system parameters. 
As shown later, the resulting evolution equation 
indicates, beyond metallic regime, the presence of eigenvalue 
correlations stronger than 
those suggested by the DMPK equation. This would lead to new  theoretical 
predictions of the statistical behavior of transport properties. 
However, in the metallic regime, the bulk eigenvalue correlations 
suggested by our evolution equation are essentially the same as those 
given by DMPK equation. 

	The paper is organized as follows. In section II, we derive the 
diffusion equation for the elements of the transfer matrix of a disordered 
region under generic scattering conditions. The equation is then used to obtain the 
statistics of transmission eigenvalues in section III. To maintain the flow of the 
discussion, only relevant steps are given in sections II, III; the details of the 
steps can be found in the appendices. The section IV deals with a derivation  of 
the "Hamiltonian" formulation of the dynamics of  transmission 
eigenvalues. The usefulness of the Hamiltonian representation is that it clearly reveals 
the hidden three body eigenvalue-interactions which are found to be absent in 
the case of DMPK equation. This is followed, in section V, by a brief discussion of 
the averages of the moments of transmission eigenvalues; the information is 
relevant for predictions of the distribution of the transport properties. The 
section VI contains a discussion about the scaling behavior of the transport properties.   
We conclude in section VII by summarizing our main results.

\section{ Statistics of Transfer Matrix Elements} 

 We consider a disordered region of length $L$ and width $W$ connected to
 two electron reservoirs by ideal (perfectly conducting) leads. The scattering
 in the region as well as reservoirs is assumed to be elastic.  The finite 
width  in the transverse direction leads to the quantization of energy of 
the transverse part of the wavefunction. As a result, the scattering states at 
the Fermi energy satisfy the relation $k^2_F=k_n^2+\epsilon_n$ with 
$k_F$ as the Fermi momentum, $k_n$ the longitudinal momentum ($k_n>0$) and $\epsilon_n$ 
as the transverse quantized eigenvalue. The various $k_n$ for $n=1,2,.., N$  
define the $N$ propagating channels. As each channel can carry two waves traveling 
in opposite directions, the wavefunction on either side of the disordered region is 
specified by a $2N$ components (corresponding to the amplitudes of $N$ 
waves propagating to the right and $N$ waves to the left).  
The normalization of the wavefunction is chosen such that it carries unit current. 
Various length-scales associated with the wavefunction divide the transport properties 
into three main regimes: (i) ballistic limit 
described by $l>L$ with $l$ as the mean free path, (ii) diffusive limit given 
by $l<L< \zeta$ with $\zeta=Nl$ as the localization length and (iii) insulator 
limit with $L>\zeta$.

 	 The scattering properties of the disordered region are completely
 characterized by a $2N \times 2N$ transfer matrix $M$ with 
$M_{kl} \equiv \sum_{s=1}^{2} (i)^{s-1} M_{kl;s}$; the subscript $s$ 
refers to the real ($s=1$) and imaginary component ($s=2$) of the element. 
The transfer matrix has a multiplicative property: if the disordered system is 
described as a sequence of $n$ "scattering units" (thin slices), with transfer 
matrices $M_1, M_2, ..., M_n$, respectively, the transfer matrix of the disordered 
system is 
\begin{eqnarray}
M^{(n)}= M_n...M_2 M_1
\end{eqnarray}

The current conservation imposes a "pseudo-unitarity" constraint on $M$:
\begin{eqnarray}
\Sigma_z M^{-1} \Sigma_z = M^{\dagger}
\end{eqnarray}
where $\Sigma_z$ is a $2N\times 2N$ matrix: 
$\Sigma_z=\pmatrix{ 1 & 0\cr \\ 0 & -1}$ with $1$ and $0$ as  
$N\times N$ unit matrix and null matrix, respectively. 
The presence of time-reversal symmetry in the disordered region 
 subjects $M$ to an additional requirement: 
\begin{eqnarray}
\Sigma_x M \Sigma_x = M^*
\end{eqnarray}
with $\Sigma_x$ as a $2N\times 2N$ matrix: 
$\Sigma_x=\pmatrix{ 0 & 1\cr \\ 1 & 0}$. 

The current conservation condition on $M$ leads to correlations between its matrix elements. 
As a result, the number of independent elements of $M$  is  $N_2=(2N)^2$. 
The time-reversal symmetry along with current conservation 
further reduces the number of independent elements to $N_1=N(2N+1)$ 
\cite{mpk,been1}. In this section, 
we denote an independent matrix element of $M$ by $M_{\mu}$ with 
$\mu$ as a single index, running from $1 \rightarrow N_{\beta}$, and replacing 
three indices $(k,l,s)$, that is, $\mu \equiv \{kl;s\}$; 
the subscript $(kl;s)$ will be used only if required for clarification.
Here $N_{\beta}=N(2\beta N+2-\beta)$ with $\beta=1$ in presence of time-reversal 
symmetry and $\beta=2$ in its absence.

	Our objective in this paper is to study the statistics of transmission 
      eigenvalues. In principle, the eigenvalues $T_n$ of the transmission 
matrix $T$ of a region are known as transmission eigenvalues. However, due to a simple 
one to one relationship with $T_n$, the doubly degenerate eigenvalues 
$\lambda_n$ of the matrix $B=[A+A^{-1}-2]/4$, with $A=M.M^{\dagger}$, are also referred as 
transmission eigenvalues: $T_n={1\over 1+\lambda_n}$, $n=1\rightarrow N$ \cite{been}. 
The $\lambda_n$ can further be expressed in terms of 
$2 N$ eigenvalues of the matrix $A$  which exist in inverse pairs and
can be denoted by $x_n$ ($n=1,2,...,2N$) \cite{been}: 
$\lambda_n=(x_n +x_n^{-1}-2)/4 $. The statistics of transmission eigenvalues 
can then be determined by a knowledge of the statistics of eigenvalues $x_n$ 
which in turn is related to the statistics of transfer matrices.

\subsection{Probability density of transfer matrices and its dependence on various 
system parameters}  
 
	For statistical analysis of transfer matrices, we consider an ensemble of 
matrices $M$, representing a collection of disordered conductors of length $L$ and 
defined by a differential probability ${\rm d}\rho(M)=\rho(M) {\rm d}\mu(M)$ 
\cite{been1,mpk,md}. Here $\rho(M)$ and ${\rm d}\mu(M)$ are the probability density 
and the invariant measure associated with the transfer matrix space (see \cite{md} 
for the details about ${\rm d}\mu(M)$). The behavior of $\rho(M)$ depends on the 
physical properties of the disordered region. This can be explained as follows:
The elements of $M$ describe the overlap between various channel states 
on two sides of the region. The presence of disorder and deterministic uncertainty 
due to complexity of region leads to randomization of the elements $M_{\mu}$. 
Based on complex nature of the disordered region, the randomness associated 
with $M_{\mu}$ can be of various types. Further the genetic scattering conditions 
can cause multiple channel interactions, resulting in varying degree of 
correlations between matrix elements.  The multiplicative property (eq.(1)) 
of the transfer matrices also imposes constraints on the behavior of $\rho(M)$: 
if $\rho_L(M)$ and $\rho_{L_0}(M_0)$ be the densities of  transfer matrices 
$M, M_0$ of a region with lengths $L, L_0$, respectively, the density $\rho_{L'}(M')$ 
of the matrix $M'=M M_0$ corresponding to the length $L'=L+L_0$ 
should be reproducible under convolutions of $\rho_L(M)$ and $\rho_{L_0}(M_0)$, 
\begin{eqnarray}
\rho_{L'}(M') = \int \rho_{L}(M' M_0^{-1}) \rho_{L_0} (M_0) {\rm d}\mu(M_0)
\end{eqnarray}
The above formulation was used as a basis to derive the DMPK equation, by 
applying it to obtain the probability density for a length $L'=L+\delta L$ with 
$L_0=\delta L$ as a small length increment. The $\rho_{\delta L} (M_0)$ 
in this case was assumed to be a maximum entropy distribution obtained under isotropy
constraints: 
$\rho_{\delta L} (M_0)= {\rm exp}(\mu-\nu {\rm Tr}M^{\dagger} M)$ with $\mu$ and 
$\nu$ as Lagrange Multipliers \cite{mpk,md}. The assumption implies that, 
in DMPK case, the information about matrix-element correlations is contained 
only in the invariant measure ${\rm d}\mu(M)$; the latter, however, contains 
only those correlations which are required to preserve the current 
conservation \cite{md}. 

	The generic scattering conditions in a disordered region result in   
the matrix element correlations beyond those due to current conservation condition.   
We therefore need to consider a probability density $\rho (M)$, free of isotropy 
constraints, with multi-channel correlations and dependent 
on various other system parameters besides length; (now onwards, we suppress 
the subscript length of $\rho$). 
In absence of any further information about the disordered region, 
the simplest and least biased hypotheses is that the system is described by the 
distribution $\rho(M)$ that maximizes Shannon's information entropy

\begin{eqnarray}
S[\rho(M)] = - \int \rho(M)\; {\rm ln} \rho(M)\; {\rm d}\mu(M)
\end{eqnarray}
under the constraints 
(i) $\rho(M)$ is normalized,  
(ii) the mean $<M_{\mu}>$ and correlations $<M_{\mu} M_{\mu'}>$ are fixed:  

$<M_{\mu}> = {\beta\over 2}{\partial {\rm log}C\over \partial a_{\mu}}$, 
$<M_{\mu} M_{\mu'}> = \beta{\partial {\rm log}C\over \partial b_{\mu\mu'}}$,  
with $C$ as the normalization constant and the parameters $a_{\mu}$, $b_{\mu\mu'}$ 
given by the system conditions. The entropy $S$ subject to above constraints can be
 maximized by considering the functional 
${\tilde S}[\rho]=S[\rho]-(1/\beta)\sum_{\mu,\mu'} b_{\mu\mu'} <M_{\mu} M_{\mu'}>  
- (2/\beta) \sum_{\mu} a_{\mu} <M_{\mu}> - C$ 
and putting its functional derivative to zero : 
\begin{eqnarray}
\delta {\tilde S} = - \int {\rm d}\mu(M)\; 
{\delta \rho(M)}\left[1+ {\rm ln} \rho(M) 
 - {1\over \beta} \sum_{\mu,\mu'} b_{\mu\mu'}  M_{\mu} M_{\mu'} - 
{2\over \beta} \sum_{\mu} a_{\mu} M_{\mu} 
\right] = 0.
\end{eqnarray}
 The above implies  
\begin{eqnarray}
\rho(M,a,b) = C{\rm exp}\left[- (1/\beta)\sum_{\mu,\mu'}
 b_{\mu\mu'}  M_{\mu} M_{\mu'} - (2/\beta) \sum_{\mu} a_{\mu} M_{\mu} \right].
\end{eqnarray}
with $C$ as a normalization constant and $a,b$ as the 
matrices of distribution parameters $a_{\mu}$ and $b_{\mu\mu'}$, respectively. 
Note the symbol $\sum_{\mu}$ implies a summation over distinct matrix elements only. 

  The Gaussian form of $\rho$ in eq.(7)  results due to consideration of constraints 
only up to second order moments of the matrix elements. The availability of information  
about higher order moments may lead to a non-Gaussian behavior of $\rho$.  
The Gaussian assumption, however, can also be justified by the same intuitive 
reasoning which led to the Gaussian distribution of $\rho(M)$ for a small length 
increment $\delta L$ in the DMPK case \cite{md}. The reasoning is based on the 
multiplicative property of $M$: Dividing the length  into small segments, the transfer 
matrix of the region can be written as a sum over matrices corresponding to the length
segments \cite{mt}. Assuming the matrices in the sum to be independent of each other,
a central limit theorem for the statistics of transfer matrices  was proved 
in \cite{mt} in the weak-scattering limit. For a length macroscopically small 
but containing many scattering units, as in the DMPK case, 
the central limit theorem justifies the assumption of independent Gaussian  
distributions for various $M_{\mu}$. For a macroscopically large length, however, 
the correlations among $M_{\mu}$'s should be taken into account which intuitively 
leads us to eq.(7). 

The distribution parameters $a_{\mu}$ and $b_{\mu\mu'}$, being measures of the 
averages of the matrix elements and their correlations, 
are influenced by various system conditions. For example, the increase of disorder 
in the region leads to localization of waves and reduced channel-channel interaction 
which in turn affects the distribution of each $M_{\mu}$. 
Similarly the increasing dimensionality $d$ of the region increases the total  number 
$N$ of existing channels, $N\approx (k_F W)^{d-1}$, as well as the probability of 
interacting channels. The latter increases the number of finite $b$ parameters. 
The boundary conditions or topology of the region also 
affect the distribution parameters, due to their influence on the interactions among
channels close to boundaries. As an example, consider the case of interaction 
between  nearest-neighbors channels only. This gives only $2N(t+1)$ finite $b$'s,  
rest of them being infinite; here $t$, the number of nearest neighbors depends 
on the dimensionality as well as boundary conditions of the region. Further the 
strength of finite $b$ parameters depends on the localization length $\zeta$ 
of the region which, in turn, is quite sensitive to the dimensionality.    

Another important system parameter affecting the sets $a,b$, is the system 
length $L$. The dependence of $a,b$ on $L$ can be derived by using the 
convolution property, given by eq.(4), as a condition. 
For example, the $L$-dependence of $b$'s can be seen by considering a simple case 
with $a_{\mu}$ length-independent. The eq.(4) gives, by using eq.(7) for  
$\rho(M')$ as well as $\rho(M)$ and writing 
$b(L+\delta L)\approx b(L)+{\partial b\over \partial L} \delta L$,
 \begin{eqnarray}
 C_{L+\delta L}\; {\rm e}^{- (1/\beta)\delta L \sum_{\mu,\mu'} 
 {\partial b_{\mu\mu'} \over \partial L} M'_{\mu} M'_{\mu'}}
 = C_{L} \int {\rm e}^{- (1/\beta) \sum_{\mu,\mu'} 
 b_{\mu\mu'} [M_{\mu} M_{\mu'} - M'_{\mu} M'_{\mu'}]} 
\rho_{\delta L} (M_0) \; {\rm d}\mu(M_0)
\end{eqnarray}
where $M=M'M_0^{-1}=M'\Sigma_z M_0^{\dagger} \Sigma_z$. The 
${\partial b_{\mu\mu'}\over \partial L}$ can now be obtained by 
multiplying eq.(8) by $\prod_{\mu} {\rm d}M'_{\mu}$, followed by an integration:
\begin{eqnarray}
{\rm Det} \left[ {\partial b\over \partial L} \right]
={C^2_{L+\delta L}\over C^2_L f^2} 
\end{eqnarray}
where $f$ as a function of $b$,  
$f(b)=\int f_0(M_0) \rho_{\delta L} (M_0) {\rm d}\mu(M_0)$,  
and $f_0$ as a function of $M_0$ and $b$: 
$f_0(M_0,b)=\int {\rm e}^{- (1/\beta) \sum_{\mu,\mu'} 
 b_{\mu\mu'} [M_{\mu} M_{\mu'} - M'_{\mu} M'_{\mu'}]} 
\; \prod_{\mu} {\rm d}M'_{\mu}$. 
The eqs.(8,9) indicate that the variation of $b$ with 
respect to $L$ depends on the scattering properties of the small 
length increment. 

	The physical properties such as current conservation and 
presence of time-reversal symmetry in the region also put restrictions on 
the strengths of the parameters $a,b$. The current conservation condition 
(eq.(2)) implies $\sum_{l} [<M_{kl} M^*_{sl}> - <M_{k,l+N} M^*_{s,l+N}>] = 
\delta_{ks} c_k$ (with $c_k=1$ for all $k\le N$ and $c_k=-1$ for $k>N$) which  
in turn connects various $b$'s (as $<M_{ij;s'} M_{kl;s}>= 
\beta {\partial {\rm log} C\over \partial b_{ijkl;ss'}}$).      
Similarly the time-reversal symmetry along with current conservation results in 
equality $M_{kl}=M_{k+N,l+N}^*$ \cite{mpk} which implies 
$b_{klij;ss'}= (-1)^{s-1} b_{k+N,l+N,ij;ss'} 
=(-1)^{s'-1} b_{kl,i+N,j+N;ss'}=(-1)^{s+s'-2} b_{k+N,l+N,i+N,j+N;ss'}$ and 
$a_{kl;s} = (-1)^{s-1} a_{k+N,l+N;s}$. 

In general, the different scattering conditions can give rise to 
different sets of distribution parameters $a,b$. For example, let us consider the 
situation in three main regimes: 

{\it (i) Ballistic Regime}:  
The transfer matrix in this regime is almost an unity matrix. 
The regime can therefore be modeled by an ensemble of matrices $M$ 
with a probability density given by eq.(7) where  
\begin{eqnarray}
a_{\mu} \rightarrow (l/2L) \delta_{\mu\mu_d}, \qquad, \qquad 
b_{\mu\mu'}\rightarrow (l/L) \delta_{\mu\mu'} ; 
\end{eqnarray} 
for all $ \mu,\mu',\mu_d$. Here $\mu_d\equiv (kk;s)$ with $M_{\mu_d} \equiv M_{kk;s}$ 
as a diagonal element. Note, for the case $L<<l$, the above 
parametric-strengths  correspond to an ensemble of  matrices with 
almost non-random elements; it approaches the ensemble of diagonal 
matrices in the limit $L/l \rightarrow 0$.

{\it (ii) Metallic Regime}:  
For this case,   (a) the overlapping between 
various channels is almost of the same strength, (b) the flow between 
two channels is not affected by the presence of other channels. The ensemble 
of matrices $M$ can then be modeled by eq.(7) with distribution parameters  
\begin{eqnarray} 
a_{\mu} \rightarrow 0, \qquad, \qquad 
b_{\mu\mu'}\rightarrow (\zeta/L) \delta_{\mu\mu'}
\end{eqnarray}
for all $ \mu,\mu',\mu_d$  and with $l<L<\zeta$. The above set of parameters 
result in an isotropic density 
$\rho(M) \propto {\rm exp}[-(\zeta/2 L){\rm Tr}(M.M^{\dagger})]$ 
which corresponds to a statistical behavior independent of system-details. 
The latter is in agreement with the observed behavior in the metallic regime.

{\it (iii) Insulator Regime:}
This regime corresponds to a zero net current flow across the disordered region
which implies an almost zero overlap between different channels states. 
The insulator state can therefore be modeled by the limit 
\begin{eqnarray} 
a_{\mu} \rightarrow 0, \qquad, \qquad 
b_{\mu\mu'}\rightarrow (L/\zeta) 
\end{eqnarray}
for almost all $\mu, \mu'$, and, with $L >> \zeta$. In large $L$ limit, above  
parameters result in an ensemble of transfer matrices with    
all matrix elements going to zero. 

\subsection{Single Parametric Evolution of $\rho(M)$} 

	The distribution of the transmission eigenvalues $\lambda_n$ can 
be obtained, in principle, from eq.(7) by using eq.(2.11) of \cite{md} which relates 
$M$ with a matrix $\tau(\lambda)$, with functions of $\lambda$ as its elements, 
and the unitary matrices $U,V$: $M=U\tau (\lambda) V$. An integration over matrices  
$U,V$, if possible, can then give the probability density for $\lambda$. The 
non-isotropic form of the distribution $\rho(M)$, however, makes the integration 
route very complicated. This motivates us to seek an alternative route. As 
discussed below, we reduce the technical complications by formulating the 
non-isotropic problem in the same form as that of the isotropic problem. This 
also helps in identifying a single parameter which governs the evolution of 
multi-parametric $\rho(M)$. 
   
	We proceed as follows. A perturbation of the disordered region due to 
a change in impurity structure or other system parameters perturbs the matrix 
elements $M_{\mu}$ and, consequently,  the probability density $\rho(M,a,b)$.  
Due to its Gaussian form,  a change in $\rho$ due to variation of $M_{\mu}$ 
can be well-mimicked by the change due to the distribution parameters $a, b$ 
(see appendix A for the derivation):

\begin{eqnarray}
L\rho= T \rho
\end{eqnarray}
where $L$ and $T$ are the operators in $M$-space and parametric space, respectively,
\begin{eqnarray}
 L &=&\sum_{\mu}{\partial \over \partial M_{\mu}}\left[{\beta\over 2}
 {\partial \over \partial M_{\mu}} + \gamma M_{\mu}\; \right]  \\
 T &=& \sum_{\mu,\mu'}  f_{\mu\mu'}
{\partial  \over \partial b_{\mu\mu'}} +  
\sum_{\mu}  f_{\mu}{\partial  \over \partial a_{\mu}}
\end{eqnarray}
with
\begin{eqnarray}
f_{\mu} &=& \gamma a_{\mu} - 2 \sum_{\mu'} a_{\mu'} b_{\mu\mu'}  \\
f_{\mu\mu'} &=& G_{\mu\mu'}\left[\gamma b_{\mu\mu'} 
-\sum_{\mu"} b_{\mu'\mu"} b_{\mu\mu"}\right]. 
\end{eqnarray} 
Here $G_{\mu\mu'}=1+\delta_{ik}\delta_{jl}$ if $\mu\equiv (kl;s)$, 
$\mu'\equiv (ij;s')$, and, 
$\delta_{xy}=1$ if $x=y$ and $\delta_{xy}=0$ for $x\not= y$.
The normalization constant $C$ is chosen such that 
$T C=\sum_{\mu}(\gamma+ (2\beta^{-1}a_{\mu}-b_{\mu\mu})) C$ 
(see appendix A for details). 
The parameter $\gamma$ is  arbitrary and marks the end of the transition. 
Note, in eq.(13), the derivatives of $\rho$ with respect to different matrix 
elements are independent of each other but the parametric derivatives are  
correlated. By considering the above combination of derivatives, therefore, 
we transfer the information regarding the correlations between elements 
of matrix $M$ to the elements of matrix $b$. This helps in reducing the 
$M$-space operator $L$ (eq.(14)), in the same form as the one in the 
isotropic case; as mentioned above, this reduction 
is useful for technical reasons as well as for comparisons with earlier studies. 

             Contrary to the single parametric evolution of $\rho$ in the 
isotropic case, the evolution of $\rho$ in eq.(13) is governed by a large 
number of parameters. However, as discussed in appendix B, 
seemingly multi-parametric diffusion can be reduced to a single parametric evolution 
by considering a transformation $(a,b) \rightarrow y$, with $y$ as a set of 
parameters $y_j$, $j=1,2,..,{\tilde N}$, which maps the parametric 
space operator $T$ (eq.(15)) as 
$T[a,b] \rightarrow T[y(a,b)] \equiv {\partial\over \partial y_1}$; 
Here ${\tilde N}$ is the number of $(a,b)$ parameters with finite strengths.  
As a result, eq.(13) can be written as 

\begin{eqnarray}
\sum_{\mu}{\partial \over \partial M_{\mu}}\left[{\beta\over 2}
{\partial {\rho}\over \partial M_{\mu}} + \gamma M_{\mu}\; 
{\rho}\right]
&=& {\partial {\rho} \over\partial Y}
\end{eqnarray}
where $Y\equiv y_1$. 
The eq.(18) is in a form of a standard Fokker-Planck equation with 
"particles" $M_{\mu}$ undergoing a Brownian motion in "time" $Y$.  
The evolution approaches a steady state in limit 
${\partial \rho\over \partial Y}\rightarrow 0$ or $Y\rightarrow\infty$ 
which occurs when $f_{\mu}, f_{\mu\mu'}\rightarrow 0$ or, equivalently, 
$a_{\mu} \rightarrow 0$, $b_{\mu\mu'}\rightarrow \gamma \delta_{\mu\mu'}$ and, 
therefore,  $\rho \rightarrow {\rm e}^{-(\gamma/2){\rm Tr}M M^{+}}$ 
(from eq.(7)). The latter is also the solution of eq.(18) in 
limit ${\partial \rho\over \partial Y}\rightarrow \infty$; 
the steady state limits of eq.(7) and (18) are, 
therefore, consistent with each other.

 	The parameter  $Y$, appearing in eq.(18), is a function of various 
parameters $a_{\mu}$ and $b_{\mu\mu'}$ which are governed by underlying 
complexity of the system; $Y$ can therefore be termed as the complexity parameter,

\begin{eqnarray}
Y &=& \sum_{\mu} \int {\rm d}a_{\mu}\; z_{\mu}\; X
+ \sum_{\mu,\mu'} \int {\rm d}b_{\mu\mu'}\; z_{\mu\mu'}\; X
+ {\rm constant }
\end{eqnarray}
where summation is implied over the distribution parameters with finite values only,   
and, $X=[\sum_{\mu} z_{\mu} f_{\mu} +\sum_{\mu,\mu'} z_{\mu\mu'} f_{\mu\mu'}]^{-1}$.  
Here the functions $z_{\mu}, z_{\mu\mu'}$ are arbitrarily chosen 
such that the ratio
\begin{eqnarray}
\left[\sum_{\mu} z_{\mu}\; {\rm d}a_{\mu} +
\sum_{\mu,\mu'} z_{\mu\mu'} \; {\rm d}b_{\mu\mu'} \right] X
\end{eqnarray}
is a complete differential; the details of the eqs.(19,20) are given in appendix B. 

	It is worth emphasizing here that the system information in  eq.(18) enters 
only through the parameter $Y$. The distribution parameters $a,b$ being system-dependent, 
$Y$ in turn is a function of the system parameters e.g. length, disorder, dimensionality 
and boundary conditions etc. A variation of any one of the system parameters e.g. 
length $L$ can therefore change $Y$ but, note, in general $Y \not= L$. (In other words, even if 
$L$ is the only system parameter subjected to change, the parameter governing 
the evolution is $Y$, a function of other system parameters besides $L$).  
However the case $Y \propto L$ can occur if the distribution parameters  are assumed to 
depend only on the length of the system (ignoring the dependence on other system 
parameters). For example, consider the case with 

\begin{eqnarray}
a_{\mu} \rightarrow 0, b_{\mu\mu'} \rightarrow (q L)^{-1} \delta_{\mu\mu'} \nonumber
\end{eqnarray}
for all $\mu, \mu'$ with $q$ as a constant. Using the values in eqs.(16, 17) gives 
$f_{\mu}=0$, $f_{\mu\mu}=2(qL)^{-2}(\gamma q L-1)$ and therefore 
$X^{-1}=2  N_{\beta}(qL)^{-2}[\gamma q L-1]$; here we have chosen 
$z_{\mu\mu'}=1$ as that makes the ratio (20) a complete differential. 
%
%with ${\tilde N}$ as the total number of finite $a_{\mu}$ and $b_{\mu\mu}$: 
%${\tilde N}=2 N_{\beta}$.   
%
Note the variation of length $L$ is this case changes only $N_{\beta}$ 
parameters $b_{\mu\mu'}$, leaving all other parameters unchanged. 
This, alongwith substitution of $X$, in eq.(19) gives 
\begin{eqnarray}
Y= q \int {{\rm d}L\over 2 (\gamma q L-1)} 
= -{1\over 2\gamma} {\rm log}|1- \gamma q L| + {\rm constant}.
\end{eqnarray}  
As obvious, the system length $L \sim (\gamma q)^{-1}$ gives $Y \rightarrow \infty$ and 
therefore steady state of the evolution: 
$\rho \rightarrow {\rm e}^{(-\gamma/2) {\rm Tr} M.M^{\dagger}}$.   

The dimensional sensitivity of $Y$ can also be explained by a simple example. 
Let us consider a disordered region with interactions between nearest-neighbor 
channels only, that is, by choosing 
\begin{eqnarray}
a_{\mu}= \delta_{\mu \mu_d}, \qquad  
b_{\mu\mu'}\rightarrow  r c_{\mu} \delta_{\mu\mu'}, 
\end{eqnarray}
where  $c_{\mu}=1$ if $M_{\mu}$ corresponds to an interaction 
between same channels (i.e. for $\mu=\mu_d \equiv (kk;s)$) or nearest 
neighbors $k,l$ ($\mu \equiv (kl;s)$), $c_{\mu}=0$ for all other $\mu$ values,  
(thus $\sum_{\mu} c_{\mu}=2N (t+1)$ with $t$ as the number of nearest neighbors).  
As channel-channel interaction depends on the localization length $\zeta$, it is 
reasonable to consider $r=\zeta/L$. For the above $a, b$-values, eqs.(16,17) give 
$f_{\mu}= (\gamma-2r) \delta_{\mu\mu_d}$, 
$f_{\mu\mu}= 2 r c_{\mu}(\gamma -r c_{\mu})$.  
Choosing again $z_{\mu}=z_{\mu\mu}=1$,  we get 
$X^{-1}=2N[\gamma+2\gamma r - 2(t+1) r^2$.  
Substituting the above in eq.(19) gives
\begin{eqnarray}
Y &=& 2\beta N \int {\rm d}r X \nonumber \\
&=& {\beta N\over 2\gamma(t+1)} 
{\rm ln}{2(t+1)\zeta + q_{+} L\over 2(t+1)\zeta - q_{-} L} + {\rm constant}
\end{eqnarray}  
with $q_{\pm}=\sqrt{\gamma^2 t^2 + (t+1)\gamma} \pm \gamma t$.  
 The parameter $t$, being a function of dimensionality as 
well as boundary conditions of the region, its presence in eq.(23) implies the 
dependence of $Y$  on the conditions. However the most significant dimension-dependence 
of $Y$  comes through the localization length $\zeta$ which varies with 
dimension of the region (see \cite{lr}). Further, as steady state of the evolution 
($Y\rightarrow \infty$) occurs at length scales $\zeta \sim \gamma L$, the approach 
to equilibrium is sensitive to system-dimension.

	The parametric space transformation $(a,b) \rightarrow y$ maps the 
probability density $\rho (M,a,b)$ to $\rho(M,y(a,b))$. As a result, $\rho$ 
depends on various parameters $y_j$, $j=1 \rightarrow {\tilde N}$ with 
 $\tilde N$ as the number of non-zero elements in set (a,b). However
the diffusion, generated by the operator $L$ in the matrix-space $M$, is governed 
by $Y\equiv y_1$ only; the rest of them, namely, $y_j$, $j>1$ remain 
constant during the evolution. Note it is always possible to define a transformation 
from the set $(a,b) \rightarrow y$ with $y_j$, $j>1$ as the constants of dynamics generated 
by $L$. This can be explained as follows. A matrix element, say $M_{\mu}$, 
describes how a basis state $\psi_i$ interacts with state $\psi_j$ through $M$. 
This results in dependence of the matrix element correlations 
and, thereby, of the set $(a,b)$, on the basis parameters e.g. basis indices.
As the basis is kept fixed during the evolution, the suitable 
functions of basis parameters can be chosen to play the role of $y_j$, $j>1$.
(Note a similar transformation has been used to obtain a single parametric 
evolution of multi-parametric Gaussian ensembles of Hermitian matrices; 
see \cite{ps1} for details).

	The eq.(18) forms the basis of our results obtained in next few sections. 
As clear, it is based on the Gaussian assumption for the probability density of the 
transfer matrices. The complexity of the region may however give rise to non-negligible 
higher order matrix element correlations which leads to non-Gaussian behavior  
of the $\rho(M)$. In this context, it is useful to note that the evolution of the density 
$\rho(H)$ of the Hermitian matrices $H$ is also known to be governed by an 
equation similar to eq.(18) (with $M$ replaced by $H$ and a different form of $Y$); 
the $\rho(H)$ can be Gaussian or non-Gaussian (see \cite{ps5} for the proof). 
Following similar steps as used for the non-Gaussian $\rho(H)$, the eq.(18) can also 
be proved for a non-Gaussian density of transfer matrices e.g. for $\rho(M) \propto
{\rm e}^{f(M)}$, with $f(M)$ now containing products of $r$ matrix elements, 
$r=1 \rightarrow n$. 

\section{Evolution of Transmission Eigenvalues Due to Changing Complexity}

	The single parametric distribution $P(\lambda,Y)$ of the transmission eigenvalues 
can, in principle, be obtained by substituting $M=U\tau (\lambda) V$ in the 
solution $\rho(M,Y)$ of eq.(18), followed by an integration over matrices $U,V$. 
However, again, as in the case of eq.(7), the integration route offers many 
technical difficulties. This again motivates us to use the differentiation option 
which has, as discussed later, other advantages too.  

        In this paper, we use the evolution of $M$, given by  eq.(18),  
to show a single parametric diffusion of the transmission eigenvalues.  
The steps are given as follows. The eq.(18) gives the first and second 
moments of the matrix element components $M_{kl;s}\equiv M_{\mu}$ as (appendix C) 
\begin{eqnarray}
\overline{\delta M_{kl;s}} &=& - \gamma M_{kl;s}  \delta Y, \qquad
\overline{\delta M_{kl;s} \delta M_{mn;s'}}
=  \beta \delta_{km}\delta_{ln}\delta_{ss'} \delta Y
\end{eqnarray}
with $\overline f$ implying an average of $f$ over random noise. 
(For clarification, we again refer a matrix element component 
as $M_{kl;s}$).   
The above moments can further be used to determine the moments of 
matrix elements $A_{mn}=\sum_{k=1}^{2N} M_{mk} M_{nk}^*$ (see appendix D for 
details),

\begin{eqnarray}
{\overline{\delta A_{mn}}} &=&  -2 \gamma A_{mn}  \delta Y \nonumber \\
{\overline{\delta A_{mn} \delta A_{mn}^* }} &=& 
2\left[\beta+(2-\beta)\delta_{mn}+ (4-\beta)\delta_{mn+N})\right] (A_{mm} + A_{nn}) 
\delta Y 
\end{eqnarray}
with $\beta=1$ in presence of the time-reversal symmetry and 
$\beta=2$ in its absence.

As mentioned before, the matrix $A$ has $2N$ eigenvalues $x_n$, 
$1\le n\le 2N$ which form inverse pairs. For clarification, 
let us label the eigenvalues such that $x_{n+N}\equiv x_n^{-1}$.  
Using $2^{nd}$ order perturbation theory for Hermitian matrices, 
the change $\delta x_n$ of the eigenvalues $x_n$ due to a small perturbation 
$\delta A$ of the matrix $A$ can be given as  
\begin{eqnarray}
\delta x_n = \delta A_{nn} + \sum_{m\not=n} 
{|\delta A_{mn}|^2 \over x_n-x_m} +O((\delta A_{mn})^3)\nonumber 
\end{eqnarray}
The above equation can then be used to obtain the first and second 
moments of the eigenvalues $x_n$, $n=1 \rightarrow N$

\begin{eqnarray}
\overline{\delta x_n} = F_n(x) \delta Y, \qquad 
\overline{\delta x_n \delta x_m } = 8 x_n \delta_{nm} \delta Y 
\end{eqnarray}
with $F_n=- 2\gamma x_n + 2\sum_{m \not= n} [\beta+(4-\beta)\delta_{m, n+N}]
(x_n+x_m)(x_n-x_m)^{-1}$ (appendix E). The eq.(26) along with the relations 
$\lambda_n = {(x_n-1)^2\over 4 x_n}$, 
$\delta \lambda_n = {(x_n^2-1)\over 4 x_n^2} \delta x_n$
gives the first and second moments of the transmission 
eigenvalues $\lambda_n$ (appendix E),

\begin{eqnarray}
\overline{\delta \lambda_n \delta \lambda_m } &=& 8  
{\lambda_n (\lambda_n+1)\over x_n} \delta_{mn} \delta Y, \qquad  
\overline{\delta \lambda_n}  = -4 E_n (\lambda_n)  \delta Y
\end{eqnarray}
where 
\begin{eqnarray}
E_n(\lambda_n)={1\over  x_n}
[{\gamma\over 2} x_n{\sqrt{\lambda_n (1+\lambda_n)}} 
-(2\lambda_n+1)
-\beta \sum_{m\not=n} {\lambda_n (1+\lambda_n)\over
\lambda_n-\lambda_m}]
\end{eqnarray}
with $x_n=1+2\lambda_n +2\sqrt{\lambda_n(\lambda_n+1)}$. The higher order 
moments of $\delta \lambda_n$ vanish to first order in $\delta Y$.

	The information about moments of the eigenvalues $\lambda_n$ can 
now be used to obtain their evolution equation. The theory of Brownian 
motion \cite{vk} informs us that the joint probability distribution 
$P(\{\lambda\})$ for the eigenvalues $\lambda_n$ evolves with increasing 
$Y$ according to Fokker-Planck equation,

\begin{eqnarray}
{\partial P \over\partial Y} &=& {1\over \delta Y}
\sum_{n=1}^N{\partial \over \partial \lambda_n}\left[{1\over 2}
{\partial (\overline{\delta\lambda_n\delta\lambda_m}) P\over \partial \lambda_n} +
(-\overline{\delta\lambda_n})\; P \right] \nonumber \\
&=& 4 \sum_{n=1}^N
{\partial \over \partial \lambda_n}\left[
{\partial (x_n^{-1}\lambda_n(1+\lambda_n) P)\over \partial \lambda_n} +
E_n \; P \right]
\end{eqnarray}

	Equation (29) describes the evolution of the transmission eigenvalues 
for a disordered region with respect to its changing complexity \cite{vk1}. 
 The reasons for a change in complexity can be manifold, for example, due to change  
in length or disorder of the region, the scattering and boundary conditions. 
As a result, eq.(29) is not only valid for higher dimensions and beyond 
weak scattering limits, it can also describe the influence of various system 
parameters on transport; the information about all of them enters in the 
probability distribution through a single parameter $Y$. Note $Y$ can also be 
expressed as a function of the localization length $\zeta$ measured in units of 
system length $L$. This can be explained as follows. 
A single parametric formulation of the diffusion in the matrix space $M$ 
implies a same parametric dependence of the evolutions in the eigenvalue 
as well as eigenfunction space. This allows a complexity parameter 
formulation of  the eigenfunction correlations  
e.g $<\psi_n (r) \psi_n(r')>=f(r,r';Y)$ (with $\psi_n(r)$ as $n^{th}$ 
eigenstate of the transfer matrix at space point $r$ of the disordered region). 
As the localization  length $\zeta$ can be expressed in terms of eigenfunction correlation 
$f(r,r'Y)$, this results in $Y$ dependence of $\zeta$.

	It is worth comparing eq.(29) with DMPK equation. The latter describes 
the evolution of transmission eigenvalues due to changing length of the 
region in the diffusive regime and for quasi one-dimensional systems.
The DMPK equation has been obtained  by assuming the elements of the transfer 
matrix of a small length $L$ of the region as independently distributed: 
$\rho_{\rm DMPK}(M) \propto {\rm e}^{-(\alpha l/2 L) {\rm Tr} M M^{\dagger}}$ with 
$\alpha=\beta N+2-\beta$. Note that $\rho_{\rm DMPK}$ is a special limit of 
eq.(7), equivalent to a choice of all $a_{\mu} \rightarrow 0$, $b_{\mu\mu'} 
\rightarrow (\alpha l/L) \delta_{\mu\mu'}$. 
Consequently the complexity  parameter $Y$ for $\rho_{\rm DMPK}$ 
turns out to be same as eq.(21) with $q=(\alpha l)^{-1}$.  In the limit $L< N l$,  
$Y$ reduces in a form similar to the evolution parameter of DMPK equation: 
$Y\rightarrow L/2\alpha l$. However the right side of eq.(29) is still 
slightly different from that of the DMPK equation. By further considering 
the strongly transmitting channels limit, namely, $\lambda_n << 1$ 
or $x_n \approx 1$, eq.(29) reduces to DMPK equation 

\begin{eqnarray}
{\partial P \over\partial Y} &=& 4 \sum_{n=1}^N
{\partial \over \partial \lambda_n}\left[
{\partial [\lambda_n(1+\lambda_n)P ]\over \partial \lambda_n} - 
(2\lambda_n+1)P - \beta \sum_{m\not=n} {\lambda_n (1+\lambda_n)P\over
\lambda_n-\lambda_m}\right]
\end{eqnarray}
Thus, for strongly transmitting channels and with length as the only changing 
parameter, eq.(29) describes the same statistical properties of the 
transmission eigenvalues as the DMPK equation. However, as eq.(30) indicates, 
if the system parameters other than length are subjected to change, 
eq.(29) has a different evolution parameter even in strongly transmitting 
channels limit although its  $\lambda$-dependent terms remain same as in the DMPK 
equation. The latter similarity is useful for the following reason.   
As many results for the length-dependence of transport properties e.g. 
conductance using DMPK equation have already been derived, a replacement 
of length by $Y$ in the known results can then give us the dependence 
of properties on other system parameters too 
(in limit $\lambda_n<<1$ only).  

\section{Hamiltonian Formulation of the Eigenvalue-Dynamics} 

	The distribution of transmission eigenvalues and their correlations 
can be used to determine the statistical behavior of various transport 
properties. For example, the conductance $G$ is given by 
$G=\sum_n (1+\lambda_n)^{-1}$ in the Landauer formulation 
(in units of dimensionless conductance $G_0=2e^2/h$) \cite{been};
consequently, its distribution is related to that of the transmission eigenvalues:
\begin{eqnarray}
P_G(G)=\int \delta(G-\sum_{n=1}^N {1 \over 1+\lambda_n}) P(\{\lambda\})
\prod_{j=1}^N {\rm d}\lambda_j
\end{eqnarray}
 A knowledge of the solution of eq.(29) is therefore very desirable.  
Although the solution is not known so far, still we can gain a deeper 
insight about eigenvalue correlations by writing eq.(29) in a 
Schrodinger equation form. 
 The required steps are as follows. A transformation of 
the eigenvalues $\lambda_n={\rm sinh}^2({\rm log}\mu_n)$ 
reduces  eq.(29) to a Fokker-Planck equation with constant 
diffusion coefficient,

\begin{eqnarray}
{\partial P \over\partial Y} &=&  \sum_{n=1}^N
{\partial \over \partial \mu_n}\left[{\partial P\over \partial \mu_n} +
\beta P {\partial \Omega(\mu_n) \over \partial \mu_n}
\right]
\end{eqnarray}
where $\Omega({\mu_n})= -\sum_{n>m}{\rm log}|\Delta(r_n,r_m)| 
- \beta^{-1} \sum_n 
[{\rm log}({\rm sinh} 2r_n) - \gamma \mu_n {\rm sinh}r_n+  r_n ]+(N(N-1){\rm ln}2)/2$ 
with $\Delta(r_n,r_m)=({\rm cosh}2r_n-{\rm cosh}2r_m)$ and $r_n={\rm log}\mu_n$.
By using $P(\{\mu\},Y)={\rm e}^{-\beta \Omega/2} \psi(\{\mu\},Y)$,
 eq.(32) can  be mapped to a Schrodinger equation 
\begin{eqnarray}
{\partial \psi \over\partial Y}= \tilde{H} \psi
\end{eqnarray}
with $\tilde H$ as the "Hamiltonian" governing the eigenvalue dynamics: 
$\tilde{H}(\{\mu\}) = \sum_n {\partial^2  \over\partial \mu_n^2} + V(\{\mu\})$
where
\begin{eqnarray}
V(\{\mu\}) &=& - {\beta\over 2}\sum_n \left[{\beta\over 2} 
\left({\partial \Omega \over\partial \mu_n}\right)^2 - 
{\partial^2 \Omega \over\partial \mu_n^2}\right] \\
&=& \sum_n {1\over \mu_n^2}\left[{1\over 4}
(2\gamma\mu_n^2-\gamma^2\mu_n^4-3)+\gamma\mu_n^2\;{\rm coth} 2r_n +
{1\over {\rm sinh}^2 2r_n}+
 \sum_m U_{mn}\right], \nonumber \\
U_{mn} &=& \left[ \beta(4+\gamma\mu_n^2) 
 {{\rm sinh}2r_n \over \Delta(r_n,r_m)}
-\beta(\beta-2){{\rm sinh}^2 2r_n \over \Delta^2(r_n,r_m)} 
-\beta^2\sum_{j,j\not=m,n} {{\rm sinh}^2 2r_n \over 
\Delta(r_n,r_m)\Delta(r_n,r_j)}\right]
\end{eqnarray}
The Schrodinger equation formulation of eq.(29) reveals an 
important aspect of the eigenvalues, namely, the presence of a three body 
interaction terms; (note it could have been converted 
into a two body term in absence of the prefactor $\mu_n^2$). 
It is well-known that the "Hamiltonian" 
appearing in the Schrodinger equation formulation of the DMPK equation 
contains only two body interaction terms \cite{been,been2}. 
Thus it appears that the presence of significant multiple channel 
interactions leads to three body eigenvalue correlations while the isotropic channel limit 
(used in DMPK case) restricts it to pairs of eigenvalues.  
The different "Hamiltonian" in the two cases result in 
a significant difference in their states $\psi$ and therefore $P(\{\mu\})$. 
As a result, the theoretical prediction of the physical properties 
under multiple channel interactions are expected to differ 
significantly from the predictions based on isotropic   
channel considerations; the suggestion is in accordance with the 
already known failure of DMPK equation beyond metallic regime or 
for two and higher dimensions.

\section{Evolution of Averages}
 
	The presence of many body interactions among eigenvalues makes it 
difficult to calculate $P(\{\lambda\})$ under general scattering conditions. 
In the metallic regime, however, the fluctuations of transport properties 
are much weaker than the average behavior and a knowledge of their averages 
(instead of full distribution) is sufficient. The evolution of transmission 
eigenvalues can then be used to study the effect of changing complexity on 
the average behavior of their various functions. Consider a function  
$F(\{\lambda\})$ not explicitly dependent on the parameter $Y$. The 
multiplication of $F$ on both sides of eq.(29), 
followed by partial integration, leads to following behavior of 
$<F>=\int F(\{\lambda\}) P(\{\lambda\}) {\rm d}\{\lambda\}$ 
with $<.>$ implying an ensemble average:
\begin{eqnarray}
{\partial <F> \over\partial Y} &=& 4\left<\sum_{n=1}^N
\left [ \lambda_n (1+\lambda_n) x_n^{-1} {\partial^2 F\over \partial \lambda_n^2}
- E_n {\partial F\over \partial \lambda_n} \right]\right>
\end{eqnarray}

 For example, evolution of the moment  $T_1^p=[\sum_n (1+\lambda_n)^{-1}]^p$, 
due to changing complexity in the strongly transmitting channel limit, can be obtained by 
 substituting $F=T_1^p$ in eq.(36) and using the limit $\lambda_n <<1$ :

\begin{eqnarray}
{\partial <T_1^p> \over\partial Y} &=& -2 p \left[\beta <T_1^{p+1}>
- (\beta-2) <T_1^{p-1}T_2> - 2(p-1) <(T_2-T_3) T_1^{p-2}> \right]
\end{eqnarray}

where notation $T_x^y$ implies $y^{\rm th}$ power of the moment 
$T_x=\sum_n (1+\lambda_n)^{-x} $. 
A substitution of $p=1, 2$ in eq.(37) then gives  
the complexity dependence of the average conductance $<G>=<T_1>$ and its 
variance $\sigma^2=<T_1^2>-<T_1>^2$, respectively,

\begin{eqnarray}
{\partial <G> \over\partial Y} &=& 
-2\left[\beta <G^2>-(\beta-2)<T_2>\right] \\
{\partial \sigma^2 \over\partial Y} &=& 
-4 \left[ \beta (<G^3>-<G><G^2>)
- (\beta-2)(<M_2><G>+<M_2 G>) -2<(T_2-T_3)> \right]
\end{eqnarray}

In general, the studies of fluctuations in transport properties about their 
average behavior require a knowledge of the averages of various 
combinations of moments e.g. $<T_u^p T_v^q>$. A substitution of 
$F=T_u^p T_v^q$ in eq.(36) leads to a hierarchy of 
coupled set of equations for the average behavior of the moments.
The details of these calculations with DMPK equation as a basis are given 
in appendix B of \cite{md};  the replacement of the length parameter 
by complexity parameter in eq.(B8) of \cite{md}, gives us the evolution of 
$<T^p T_q^r>$ with respect to changing complexity of a system in strongly 
transmitting channel limit.    

 The coupled form of eqs.(38,39) or eq.(B8) of \cite{md} 
makes it difficult to obtain the 
exact behavior of the moments, even in $\lambda_n<<1$ limit . However, by 
assuming the number $N$ of the channels very large, the equations can be 
solved by using $N^{-1}$ expansion of the moments, see \cite{md} for details.  
For example, the evolution of $T_1^p$ reduces to, in the 
leading order of $N$,    

\begin{eqnarray}
{\partial <T_1^p> \over\partial Y} &=&  
- 2 \beta p  \left< T_1^{p+1}\right> + o(N^{p})
\end{eqnarray}
In ballistic limit, the eigenvalues $\lambda_n \rightarrow 0$ 
(for $n=1\rightarrow N$) 
which gives $<T_1^p>=N^p$. For ballistic case as the initial state of the 
disordered region, that is,  ${\rm lim}\; Y \rightarrow Y_0 <T_1^p>=N^p$, 
the solution of eq.(40) can be given as 
\begin{eqnarray}
<T_1^p>=N^p (1+\Lambda)^{-p} + o(N^{p-1})
\end{eqnarray}
 with $\Lambda=2\beta N|Y-Y_0|$. Following eq.(21), $\Lambda=q L/2$ 
for almost same strength of interactions between any two channels and 
length as the only changing parameter (for $\gamma q L<1$). The behavior of $<T_1^p>$ 
with respect to the above $\Lambda$  coincides with one obtained by using 
DMPK equation for which $q^{-1}=(\beta N+\beta-2) l$ (see eq.(3.31) of \cite{been}).  
The substitution of 
$p=1$ in eq.(41) gives  $Y$ dependence of the average conductance 
$<G>=<T_1>\approx N\Lambda^{-1} +o(N^0)$ (in large $N$ limit). 
The average conductance of a wire therefore decreases linearly 
with increasing length which is in agreement with the already known 
results \cite{been}.

     Equations (22,23) can be used to study the effect of changing complexity 
on the conductance fluctuations in the diffusive regime. 
For example, the weak localization correction $\delta G=<G>-(1+\Lambda)^{-1}$ 
to the average conductance as well as the variance 
$\sigma^2=<G^2>-<G>^2$ can be obtained by expanding the moments 
 to order $N^{-1}$:
\begin{eqnarray}
\delta G &=& (1-2/\beta)\Lambda^3/3(1+\Lambda)^{3} + o(N^{-1}) \\
{\rm Var}\; G &=& (2/15\beta)[1-(1+6\Lambda)/(1+\Lambda)^{6}] + o(N^{-1}).
\end{eqnarray}
(Above results were calculated in \cite{md} for length variation. Due to  
similarity of eq.(30) with DMPK equation in limit $\lambda_n <<1$, 
 same results can be used to describe the complexity dependence of conductance 
fluctuations just by replacing $L/l$ by $\Lambda$).    
The diffusive limit $\Lambda \rightarrow \infty$ of the above results give 
$\delta G=(1-2/\beta)/3$ and ${\rm Var}\; G=(2/15\beta)$ which is in 
agreement with expected universal behavior of the conductance fluctuations 
in this limit (see \cite{been} for the latter).

As mentioned above, eqs.(37-43) are valid only in the limit $\lambda_n << 1$. 
A detailed analysis of the dependence of transport properties on $\Lambda$,  
beyond strongly transmitting channel limit, is still under investigation; 
the results will be given elsewhere.

\section{Single Parameter Scaling of Transport Properties}

	A distribution $P(X)$ of a physical quantity $X$ that depends on system 
size $N$ and a set of system parameters $\{alpha_k\}$ obeys one parameter scaling 
if it is a function of only $X$ and one scale dependent parameter: 
$P(X;L;\{\alpha_k\}) = f(X;Y)$ \cite{mj}. Thus the system-information appears in the 
distribution only through the scaling parameter $Y\equiv Y(N,\{\alpha_k\})$. 

Following the discussion given in sections II, III, it is clear that the 
distribution of transmission eigenvalues satisfies the above requirement, that is,  
$P(\{\lambda_n\}; N;a,b)=P(\{\lambda_n\};Y)$, with complexity 
parameter $Y=Y(N,a,b)$ playing the role of scaling parameter. As a consequence, the 
distribution $P_F(F)$ of any transport property $F(\{\lambda\})$, 
would also follow a single parameter scaling: 
$P_F(F;Y)=\int \delta(F-F(\{\lambda\})) P(\{\lambda\};Y)\prod_{j=1}^N {\rm d}\lambda_j$

	The $Y$ governed evolution of the distribution $P_F(F;Y)$ can be obtained 
with the help of eq.(29). The evolution as a function of $Y$, however, is abrupt in 
large $N$-limit (see, for example, eqs.(40)); a smooth crossover can be seen only 
in terms of a rescaled parameter $\Lambda$ where  
\begin{eqnarray}
\Lambda={Y-Y_0\over D}
\end{eqnarray} 
with $D$ as the local mean level spacing of transmission eigenvalues $\lambda_n$. 
(Note the relation $\Lambda=2\beta N|Y-Y_0|$, given in section V, is valid only in the 
strongly transmitting channel regime).  

	The parameter $\Lambda$ governs the statistical behavior of the 
transport properties.
As discussed in section I, the dimensional-sensitivity of the 
interaction between various channels can influence  both $Y$ and $D$. This 
in turn leads to dimensional-dependence of $\Lambda$. The 
statistical behavior of transport properties e.g. conductance is therefore 
expected to be different for higher dimensions. The size-dependence of transport 
properties and influence of scattering conditions on them can also be explained 
by $Y$ and therefore $\Lambda$-dependence on these system conditions. 
 
 The one-parameter scaling behavior of the distribution $P(\{\lambda\})$ implies 
the existence of a universal distribution 
$P^*(\{\lambda\})= {\matrix {{\rm lim} \cr \\{N\rightarrow \infty}}} 
P(\{\lambda\},\Lambda)$ at a critical point which is fixed by the critical value 
$\Lambda^*={\matrix {{\rm lim} \cr \\{N\rightarrow \infty}}} \Lambda (N)$.    
Thus the size-dependence of $\Lambda$ plays a crucial role in locating the critical 
point of statistics.  It is possible that various system-conditions in a disordered 
region may result in different $N$-dependence of $|Y-Y_0|$ 
and $D$, say, $|Y-Y_0| \propto N^{\alpha}$ and $D \propto N^{\eta} $ which gives  
$\Lambda \propto N^{\alpha-\eta}$.
In regions connected with leads of finite width (implying finite 
$N$) , a variation of size $N$, therefore, leads to a smooth crossover of 
statistics  between an initial state ($\Lambda \rightarrow 0$) and the equilibrium  
($\Lambda \rightarrow \infty$); the intermediate statistics belongs to an infinite 
family of ensembles, parameterized by $\Lambda$ and given by the solution of eq.(29). 
However, for system-conditions leading to $\alpha=\eta$, the statistics 
becomes universal for all sizes, $\Lambda$ being $N$-independent; the corresponding 
system conditions can then be referred as the critical conditions.     

As clear from above, a variation of system conditions (i.e. changing $\alpha, \eta$) 
for a given finite system size, would  also lead to a crossover of statistics. 
In large $N$ limit, however, the variation causes a transition of statistics.  
Based on relative values of $\alpha$ and $\eta$, the statistics for large system sizes   
can be divided into three different types

{\it Case $\alpha < \eta$}: This corresponds to $\Lambda \rightarrow 0$ in 
limit $N\rightarrow \infty$. The system under 
variation of such conditions rapidly approaches  its initial state. 

{\it Case $\alpha > \eta$}: This gives $\Lambda \rightarrow \infty$ for 
infinite system sizes; as a result,  even a small perturbation of the system 
will cause an abrupt transition from the initial state to the equilibrium. 

{\it Case $\alpha=\beta$}: Here $\Lambda$ being size-independent, the 
case corresponds to the critical point of evolution. Further, due to finite  
and non-zero $\Lambda$,  the eigenvalue statistics 
for this case is different, even in limit $N\rightarrow \infty$, 
 from both, the initial state as well as the equilibrium. Note, however, the 
statistics is same as the one at the critical point of finite systems.

As clear from preceding sections (III-V), the $\Lambda$-formulation of transport 
properties (e.g. eqs.(29-43)) is based on eq.(18). Although this paper 
contains the derivation of eq.(18) only for Gaussian $\rho(M)$, 
but, for the reasons mentioned in section II, we expect it  
to be valid for non-Gaussian cases too. This suggests the application of 
our results for a wide range of disordered regions.  
Also note that eq.(18) and eq.(29) are valid under inelastic 
scattering conditions too. However the connection between the transport 
properties and $\lambda$'s needs to be revised. For example, the Landauer's 
formula $G=\sum_n (1+\lambda_n)^{-1}$ is not applicable in the inelastic regime;  
the results for Conductance fluctuations given in section $V$ are therefore not 
valid for that regime.
 
	The validity of eqs.(18,29) for non-Gaussian cases as well as inelastic 
conditions  suggests the existence of a $\Lambda$-formulation of transport properties 
for regions with generic scattering and disorder conditions. Although the 
functional form of $\Lambda$ is expected to be different under generic conditions,  
the critical statistics would still occur at system conditions satisfying the critical 
criteria $\alpha=\beta$, thus leading to an $N$-independent $\Lambda$. 
	It should be stressed, however, that the critical criteria may not be 
fulfilled for all disordered systems; such systems will not show any transition 
in the statistics of transmission eigenvalues.

\section{conclusion}
 
        In summary, we find that, under generic 
scattering conditions, the eigenvalues are subjected to many body interactions 
besides logarithmic repulsion. As a consequence, the correlations among eigenvalues 
and thereby the fluctuations of transport properties beyond metallic regime 
are significantly different from those within the regime. An explicit formulation 
of the fluctuations can  be obtained by the knowledge of the solution of eq.(29) 
(or eq.(33)). Although a complete solution is not known so far, nonetheless 
eq.(29) reveals, for the 
first time, a very important characteristic of transport phenomena through disordered 
regions, namely, their dependence on a single complexity parameter $Y$. This implies that  
the disordered regions with different system parameters e.g., boundary conditions, scattering 
conditions etc. will show same fluctuations of transport properties if their  
parameters $\Lambda$ correspond to a same value. This implies a deeper level of universality 
lying underneath the transport phenomena, even beyond metallic regime 
(where it was known to exist so far). A knowledge of universality is  useful from 
a practical viewpoint, besides other reasons. This is because the information about the 
transport properties of a simple disordered region (obtained by some other technique) 
can now be used to determine those for a complicated region, just by comparing their 
complexity parameters. Further the dependence of distribution of the transmission eigenvalues 
 on various system parameters through a single parameter $Y$ also indicates the validity 
of one parameter scaling theory of localization for statistical behavior of the conductance 
(also for those transport properties which can be formulated as the functions of 
transmission eigenvalues).     

\section{Acknowledgements}

	It is a pleasure to acknowledge the useful discussions with K.A.Muttalib and 
M.Stepanov.

\begin{appendix}

\section{Proof of equation (13)}

Using eq.(7), a derivative of $\rho_0=\rho/C$ with respect to 
$M_{\mu} \equiv M_{kl;s}$ can be written as

\begin{eqnarray}
{\partial \rho_0 \over \partial M_{\mu}} 
&=& - {2\over \beta} \left(\sum_{\mu'}b_{\mu\mu'} M_{\mu'} +a_{\mu} \right)\rho_0
\end{eqnarray}
where $b_{\mu\mu'}=b_{\mu'\mu}$. 
Further 
\begin{eqnarray} 
G_{\mu\mu'}{\partial \rho_0 \over \partial b_{\mu\mu'}} = 
-{2\over\beta}  M_{\mu} M_{\mu'} \rho_0, \\
{\partial \rho_0 \over \partial a_{\mu}} = -{2\over\beta}  M_{\mu}  \rho_0, 
\end{eqnarray} 
where $G_{\mu\mu'}=1+\delta_{ij}\delta_{kl}$ if $\mu \equiv \{kl;s\}$, 
$\mu'\equiv \{ij;s'\}$.  The eqs.(A1,A2,A3) can be combined to give 
\begin{eqnarray}
\sum_{\mu} M_{\mu} {\partial \rho_0\over \partial M_{\mu}}
&=&  \sum_{\mu,\mu'} G_{\mu\mu'} b_{\mu\mu'} 
{\partial \rho\over \partial b_{\mu\mu'}} + 
\sum_{\mu} a_{\mu} {\partial \rho_0\over \partial a_{\mu}}.
\end{eqnarray}

Similarly 
\begin{eqnarray}
{\partial^2 \rho_0 \over \partial M_{\mu}^2} &=& 
\left({2\over\beta}\right) (2 \beta^{-1} a^2_{\mu} - b_{\mu\mu})\; \rho_0 + 
{4\over\beta^2} \sum_{\mu'} b_{\mu\mu'} M_{\mu'} \left(2 a_{\mu} + \sum_{\mu"}  
b_{\mu\mu"} M_{\mu"} \right) \; \rho_0   \\
&=& \left({2\over\beta}\right)(2\beta^{-1} a^2_{\mu}-b_{\mu\mu}) \rho_0 - 
{2\over\beta}\left( \sum_{\mu',\mu"} G_{\mu'\mu"} b_{\mu\mu'} b_{\mu\mu"}
{\partial \rho_0\over \partial b_{\mu'\mu"}} + 
2\sum_{\mu'} a_{\mu} b_{\mu\mu'} {\partial \rho_0\over \partial a_{\mu'}}\right).
\end{eqnarray}
Using above equalities, 

\begin{eqnarray}
& & \sum_{\mu}{\partial \over \partial M_{\mu}}\left[{\beta\over 2}
 {\partial \rho_0\over \partial M_{\mu}} + \gamma M_{\mu}\; \rho_0\right] \\
&=& \sum_{\mu,\mu'}\left[\gamma G_{\mu\mu'} b_{\mu \mu'} 
{\partial \rho_0\over \partial b_{\mu \mu'}}
-\sum_{\mu"} G_{\mu'\mu"} b_{\mu\mu'} b_{\mu\mu"}
{\partial \rho_0\over \partial b_{\mu'\mu"}}\right] + 
\sum_{\mu} a_{\mu}\left[\gamma {\partial \rho_0\over \partial a_{\mu}} 
- 2 \sum_{\mu'} b_{\mu\mu'} {\partial \rho_0\over \partial a_{\mu'}} \right]
+ C_{\beta} \rho_0 
\end{eqnarray}
where $C_{\beta}=\sum_{\mu}\left[\gamma + (2\beta^{-1} a^2_{\mu} - b_{\mu\mu})\right]$. 
 Now  interchanging the indices in the terms inside the square brackets in 
eq.(A10) gives  
\begin{eqnarray}
\sum_{\mu}{\partial \over \partial M_{\mu}}\left[{\beta\over 2}
{\partial \rho_0\over \partial M_{\mu}} + \gamma M_{\mu}\; \rho_0\right] 
&=& \sum_{\mu,\mu'} f_{\mu\mu'} {\partial \rho_0\over \partial b_{\mu \mu'}}
+ \sum_{\mu} f_{\mu} {\partial \rho_0\over \partial a_{\mu}} 
+ C_{\beta} \rho_0
\end{eqnarray}
where $f_{\mu}$ and $f_{\mu\mu'}$ are given by eqs.(16,17). Multiplying 
eq.(A9) by $C$ and using $C\rho_0= \rho$, we get 

 \begin{eqnarray}
\sum_{\mu}{\partial \over \partial M_{\mu}}\left[{\beta\over 2}
{\partial \rho\over \partial M_{\mu}} + \gamma M_{\mu}\; \rho\right] 
&=& \sum_{\mu,\mu'} f_{\mu\mu'} \left[{\partial \rho\over \partial b_{\mu \mu'}} 
- \rho {\partial C\over \partial b_{\mu \mu'}} \right] 
+ \sum_{\mu} f_{\mu} \left[{\partial \rho\over \partial a_{\mu}} 
- \rho {\partial C\over \partial a_{\mu}}\right] + C_{\beta} \rho
\end{eqnarray} 
The eq.(13) can now be obtained by choosing the normalization condition on 
$\rho$ such that $C$ satisfies the relation 
$\sum_{\mu,\mu'} f_{\mu\mu'} {\partial C\over \partial b_{\mu \mu'}}  
+\sum_{\mu} f_{\mu} {\partial C\over \partial a_{\mu}}= C_{\beta} C$.

\section{Single Parametric Form of $T$: Derivation of Eq.(18)}

 	Consider a transformation from the ${\tilde N}$-dimensional $\{a,b\}$-space to 
another parametric space, say $y$-space, consisting of variables $y_i$,
$i=1 \rightarrow {\tilde N}$ which reduces $T(a,b)$ (eq.(15)) into 
\begin{eqnarray}
T(y[a,b]) \rho \equiv {\partial \rho \over\partial y_1} |_{y_2,..,y_M}
\end{eqnarray}. 
The conditions to determine desired transformation can be obtained as follows, 
By using partial differentiation, $T(b)$ given by eq.(15) can
 be transformed in terms of the derivatives with respect to $y$,
\begin{eqnarray}
T \rho = \sum_k t_k {\partial \rho\over \partial y_k}
\end{eqnarray}
where
\begin{eqnarray}
t_k \equiv \sum_{\mu} f_{\mu}{\partial y_k \over \partial a_{\mu}}
+ \sum_{\mu,\mu'} f_{\mu\mu'}{\partial  y_k\over \partial b_{\mu\mu'}}.
\end{eqnarray}
The eq.(B2) can be reduced in the desired form of eq.(B1), if the transformation
$b\rightarrow y$ satisfies following condition:
\begin{eqnarray}
t_k = \delta_{k1} \qquad\qquad {\rm for}\; k=1\rightarrow {\tilde N} 
\end{eqnarray}

According to the theory of partial differential equations (PDE)
\cite{sne},
the general solution of a linear PDE
\begin{eqnarray}
\sum_{i=1}^{\tilde N} P_i(x_1,x_2,..,x_{\tilde N}){\partial Z \over \partial x_i} = R
\end{eqnarray}
is $F(u_1,u_2,..,u_n)=0$ where $F$ is an arbitrary function and
$u_i(x_1,x_2,..,x_n,Z)=c_i$ (a constant), $i=1,2,..,n$ are independent
solutions of the following equation
                                                                                                                        
\begin{eqnarray}
{{\rm d}x_1 \over P_1} =
{{\rm d}x_2 \over P_2} =.....
{{\rm d}x_k \over P_k} =......
{{\rm d}x_{\tilde N} \over P_{\tilde N}} =
{{\rm d}Z \over R}
\end{eqnarray}
                                                                                                                        
Note the function $F$ being arbitrary, it can also be chosen as
\begin{eqnarray}
F\equiv\sum_j (u_j-c_j)=0
\end{eqnarray}
                                                                                                                        
        The equations for various $y_j$ in  the set of eq.(B4) are of the
same form as eq.(B5) and, therefore, can be solved as described above.
Let us first consider the equation for $y_1$; its general solution can be given
by a relation $F(u_{1},u_{2},..,u_{\tilde N}) = 0$ where  function $F$ is arbitrary 
and $u_j$ are the functions of ${\tilde N}$ parameters of set $\{a,b\}$ such 
that $u_{j}(\{a,b\},y_1)=c_j$ (with $c_j$'s as constants). 
The functions $u_{j}$ are the independent solutions of the equation
                                                                                                                        
\begin{eqnarray}
{{\rm d}b_{\mu\mu'} \over f_{\mu\mu'}} =....=
{{\rm d}a_{\mu} \over f_{\mu}} =......
={{\rm d}y_1}
\end{eqnarray}
where the equality between ratios is implied for all possible indices $\mu, \mu'$
(with the total number of ratios as ${\tilde N}+1$).
It is easy to see that each of the above ratios is equal to
${\sum_{\mu,\mu'} z_{\mu\mu'} {\rm d}b_{\mu\mu'} + \sum_{\mu} z_{\mu} {\rm d}a_{\mu} 
\over \sum_{\mu,\mu'} z_{\mu\mu'} f_{\mu\mu'} + \sum_{\mu} z_{\mu} f_{\mu}} $ 
where $z_{\mu}, z_{\mu\mu'}$ are arbitrary functions. The eq.(B8) can therefore be 
rewritten as
\begin{eqnarray}
{\rm d}y_1=
{\sum_{\mu,\mu'} z_{\mu\mu'} {\rm d}b_{\mu\mu'} + \sum_{\mu} z_{\mu} {\rm d}a_{\mu} 
\over \sum_{\mu,\mu'} z_{\mu\mu'} f_{\mu\mu'} + \sum_{\mu} z_{\mu} f_{\mu}}  
\end{eqnarray}
A solution, say $u_1$ of eq.(B9), or alternatively eq.(B8), can now be obtained by
choosing the functions in the set $z$ such that the right side of the above equation
becomes an exact differential:
\begin{eqnarray}
u_1\equiv y_1 - \sum_{\mu} \int {\rm d}a_{\mu} z_{\mu}\; X  
- \sum_{\mu,\mu'} \int {\rm d}b_{\mu\mu'} z_{\mu\mu'}\; X   = constant
\end{eqnarray}
where $X=[\sum_{\mu} z_{\mu} f_{\mu} +\sum_{\mu,\mu'} z_{\mu\mu'} f_{\mu\mu'}]^{-1}$. 
The general solution for $y_1$ can therefore be given by a combination of all
possible functions $u$ obtained by using arbitrary set of $z$-functions.
%It can be shown that only $M$ of the solutions  would be independent and
It can be shown that each such solution differs from the other
only by a constant: $u_j=u_i$ + constant; (this is due to equality of the two ratios
obtained by choosing two different sets $z$ of the functions).
 The $y_1$ can therefore be written as follows,
\begin{eqnarray}
y_1 =  \sum_{\mu} \int {\rm d}a_{\mu} \; z_{\mu}\; X + 
\sum_{\mu,\mu'} \int  {\rm d}b_{\mu\mu'} \; z_{\mu\mu'} \; X  + constant
\end{eqnarray}
which gives eq.(19) for $Y\equiv y_1$.

The set of equations (B4) can similarly be solved  for other $y_j$ ($j>2$).
For example, the solution for $y_k$ can be given by the
function $F_k(v_1,..v_M)=0$ where $v_j(\{a,b\},y_k)=constant$ are the independent
solutions of following equality
                                                                                                   
\begin{eqnarray}
{{\rm d}b_{\mu\mu'} \over f_{\mu\mu'}} =....=
{{\rm d}a_{\mu} \over f_{\mu}} =......
={{\rm d}y_k \over 0}
\end{eqnarray}
A solution, say $v_1$, of eq.(B12) can now be given as
\begin{eqnarray}
v_1\equiv y_k - \sum_{\mu,\mu'} \int {\rm d}b_{\mu\mu'} z_{\mu\mu'} 
- \sum_{\mu} \int {\rm d}a_{\mu} z_{\mu} 
= constant
\end{eqnarray}
where the set $z$ is a set of arbitrarily chosen  functions which satisfy 
the condition
\begin{eqnarray}
\sum_{\mu,\mu'} z_{\mu\mu'} f_{\mu\mu'} + \sum_{\mu} z_{\mu} f_{\mu} =0
\end{eqnarray}
As obvious, one possible choice for $z$ functions satisfying  the above condition
is $z_{\mu\mu'}=0$, $z_{\mu}=0$ for all $\mu, \mu'$ which gives $y_k=constant$.
As each solution of eq.(B12) is different from the other only by a constant,
the general solution for $y_k$, $k>1$, can now be given as
\begin{eqnarray}
y_k =  \sum_{\mu,\mu'} \int {\rm d}b_{\mu\mu'} z_{\mu\mu'} 
+\sum_{\mu} \int {\rm d}a_{\mu} z_{\mu} + {\rm constant}. 
\end{eqnarray}

\section{Proof of equation (7)}

	The eq.(7) can be obtained from eq.(18), by following standard routes 
of calculation of the moments from a Fokker-Planck equation; see \cite{meta,vk} 
for details. For completeness sake,  we describe here one such route.  

	Let $\rho(M-\delta M;Y)$ be the joint probability density that the 
elements of transfer matrix will be at the positions  
$M-\delta M\equiv \{M_{\mu}-\delta M_{\mu}\}$
at parameter value $Y$. At $Y+\delta Y$, let the positions of the 
elements change to $M\equiv \{M_{\mu}\}$. The changes $\delta M_{\mu}$,  
for all $\mu$ values, are random variables and  different, in general, for 
every member of the ensemble. The probability density $\rho(M;Y+\delta Y)$ 
can be expressed in terms of $\rho(M-\delta M;Y)$:  

\begin{eqnarray}
\rho(M;Y+\delta Y)= \int {\rm d}(\delta M)
\psi(M-\delta M,Y| M,Y+\delta Y) \rho(M-\delta M; Y) 
\end{eqnarray}
where $\psi$ is the probability that the position of the transfer matrix 
will change from $M-\delta M$ to $M$ in a parametric interval $\delta Y$. 
Expanding $\rho$ on both sides of eq.(C1) in a power series of $\delta M$ 
and $\delta Y$ around $\rho(M,Y)$, one gets

\begin{eqnarray}
& &\rho(M,Y) + {\partial \rho\over \partial Y} \delta Y + o((\delta Y)^2) 
=  \nonumber \\
& &\int {\rm d}(\delta M) \psi 
\left[\rho(M,Y) - \sum_{\mu} {\partial \rho\over \partial M_{\mu}} \delta M_{\mu} 
+ {1\over 2}\sum_{klij;ss'} {\partial^2 \rho\over \partial M_{\mu} M_{ij;s'}} 
(\delta M_{\mu} \delta M_{ij;s'}) + o((\delta M)^3\right]   
\end{eqnarray}
Now by using the definition 
$\overline{(\delta M)^n} = \int (\delta M)^n \psi(M-\delta M,Y|M,Y+\delta Y)
 {\rm d}(\delta M)$ and 
the equality $\int \psi(M-\delta M,Y|M,Y+\delta Y){\rm d}(\delta M)=1$, 
the eq.(C2) can be reduced to 

\begin{eqnarray}
 {\partial \rho\over \partial Y} \delta Y 
 = {1\over 2}\sum_{\mu,\mu'} \left[\overline{(\delta M_{\mu}\delta M_{\mu'})}\right]
{\partial^2 \rho\over \partial M_{\mu} \partial M_{\mu'}} - \sum_{\mu}
\left[\overline{\delta M_{\mu}} \right] {\partial \rho \over \partial M_{\mu}}  
\end{eqnarray}
By comparing eq.(C3) with eq.(18)  one obtains eq.(24).  The latter can also 
be derived by a direct integration of eq.(18); see \cite{meta} for details.

.
\section{Proof of equation (25)}

   The matrix $A$ is defined by $A=M.M^{+}$ with its elements  
$A_{mn}=\sum_{k=1}^{2N} M_{mk} M_{nk}^*$. The average of a small change  
$\delta A$ over random noise can then be expressed 
in terms of $\delta M$:

\begin{eqnarray}
\overline{\delta A_{mn}} 
&=&  \sum_{k=1}^{2N} \left[M_{mk} \overline{\delta M_{nk}^*}+
M_{nk}^* \overline{ \delta M_{mk}}\right] \\
&=& -\gamma\sum_{k=1}^{2N} 
\left[M_{mk} M_{nk}^*+M_{nk}^* M_{mk}\right]
= -2 \gamma A_{mn}  \delta Y.
\end{eqnarray}
Here eq.(D2) is obtained from eq.(D1) by using the relation (24).

	The current conservation condition on the transfer matrix $M$, given by eq.(2), 
introduces various relations between its matrix elements: 
\begin{eqnarray}
\sum_{k=1}^N \left[ M_{nk} M_{mk}^* - M_{n k+N} M_{m,k+N }^* \right] 
= g\delta_{nm} \\
\sum_{k=1}^N \left[ M_{nk} M_{n+N k}^* - M_{n k+N} M_{n+N k+N}^* \right] = 0 
\end{eqnarray}  
with $g=1$ if $n\le m$ and $g=-1$ for $n>m$. 
The presence of time-reversal symmetry alongwith current conservation, 
further implies that $M_{mk}=M_{|m+N| |k+N|}^*$ ($|x|=x$ 
if $x \le 2N$, $|x|=x-2N$ if $x>2N)$.   
As a result, the second moment of $\delta A_{mn}$ depends on the presence or 
absence of time-reversal symmetry in the disordered region and is different for 
the cases $m=n$, $m\not=n$ and $m=n+N$.

\subsection{Case $m\not= n$, $m\not=n+N$}

\begin{eqnarray}
{\overline{\delta A_{mn} \delta A_{mn}^* }}  
&=& \sum_{k,l=1}^{2N} {\overline{
\left[M_{mk} \delta M_{nk}^*+M_{nk}^* \delta M_{mk}\right] 
\left[M_{ml}^* \delta M_{nl}+M_{nl} \delta M_{ml}^*\right]}} \\
&=& \sum_{k,l=1}^{2N} \left[
M_{mk} M_{ml}^* {\overline {(\delta M_{nk}^* \delta M_{nl})}}
+M_{nk}^* M_{nl} {\overline {(\delta M_{mk} \delta M_{ml}^*)}} 
\right. \nonumber \\ 
&+& \left .  M_{mk} M_{nl} {\overline {(\delta M_{nk}^* \delta M_{ml}^*)}}
+M_{nk}^* M_{ml}^* {\overline {(\delta M_{mk} \delta M_{nl}})} \right ]
\end{eqnarray}
 
 Following eq.(24), we have  
 
\begin{eqnarray}
{\overline {(\delta M_{jk} \delta M_{rl})}} &=& 
\sum_{s_1,s_2}  i^{s_1+s_2-2} 
{\overline {(\delta M_{jk;s_1} \delta M_{rl;s_2})}} = 0 \nonumber \\
{\overline {(\delta M_{jk}^* \delta M_{rl})}} &=& 
\sum_{s_1,s_2} (-1)^{s_1-1} i^{s_1+s_2-2} 
{\overline {(\delta M_{jk;s_1} \delta M_{rl;s_2})}} =   
2\beta \delta_{jr} \delta_{kl}\delta Y
\end{eqnarray}
Consequently the last two terms inside the square bracket $[ ]$ in 
eq.(D6) do not contribute in this case and, with help of eq.(24), we get 

\begin{eqnarray}
\overline{\delta A_{mn} \delta A_{mn}^* }  &=& 
2 \beta \delta Y \sum_{k,l}^{2 N} \left [ 
\delta_{kl} M_{mk} M_{ml}^*+ \delta_{kl} M_{nk}^* M_{nl} \right ] \\
&=& 2\beta [A_{mm}+A_{nn}] \delta Y
\end{eqnarray}
 
\subsection{ Case $m=n+N$}

     Using current conservation condition given by eq.(D4), we can write 
\begin{eqnarray}
A_{n n+N}= 2\sum_{k=1}^N M_{nk} M_{n+N k}^*
\end{eqnarray}
The above equation gives
\begin{eqnarray}
{\overline{\delta A_{n,n+N} \delta A_{n,n+N}^* }}  
&=& 4\sum_{k,l=1}^{N} \left[
M_{n+N,k}^* M_{n+N,l}{\overline {(\delta M_{nk} \delta M_{nl}^*)}}
+M_{nk} M_{nl}^* {\overline {(\delta M_{n+N,k}^* \delta M_{n+N,l})}} 
\right. \nonumber \\ 
&+& \left .  M_{n+N,l} M_{nk} {\overline {(\delta M_{nl}^* \delta M_{n+N,k}^*)}}
+M_{nl}^* M_{n+N,k}^* {\overline {(\delta M_{n+N,l} \delta M_{nk}})} \right ]
\end{eqnarray}

In absence of time-reversal symmetry, the contribution from the last two terms 
inside the bracket is zero (see eq.(D7)). However the time-reversal symmetry 
along with current conservation condition in a region implies  
 $M_{mk}=M_{|m+N| |k+N|}^*$. As a result, the last two terms 
inside the bracket can be rewritten as follows

\begin{eqnarray}
 M_{nk} M_{n+N,l} \overline {(\delta M_{n+N,k}^* \delta M_{nl}^*)}
&=& M_{|n+N|,|k+N|}^* M_{n+N,l} \overline {(\delta M_{n|k+N|} \delta M_{nl}^*)}  
\delta_{\beta 1}
\nonumber \\
M_{n+N,k}^* M_{nl}^* \overline {(\delta M_{nk} \delta M_{n+N,l})} 
&=& M_{n,|k+N|} M_{nl}^* \overline 
{(\delta M_{n,|l+N|}^* \delta M_{nk})} \delta_{\beta 1} 
\end{eqnarray}

The  presence of time-reversal symmetry as well its absence therefore results in  

\begin{eqnarray}
\overline{|\delta A_{n,n+N}|^2 }  
&=& 8\beta \delta Y \sum_{k,l=1}^{N} \delta_{kl}
\left[ (M_{n+N,k}^* M_{n+N,l} +M_{nk} M_{nl}^*)  (1+\delta_{1\beta})  \right ]
\end{eqnarray}
The above equation can further be simplified by using eq.(D3) which gives 
$A_{jj}=2\sum_{k=1}^N M_{jk} M_{jk}^* - g$. Using the equality in eq.(), we 
get 

\begin{eqnarray}
\overline{|\delta A_{n,n+N}|^2 } &=& 8 (A_{nn}+A_{n+N,n+N})\delta Y
\end{eqnarray}

\subsection{Case $m=n$}

For a clear understanding, here we consider the cases with and without 
time-reversal symmetry separately. As shown below, 
$\overline{|\delta A_{nn}|^2}$ turn out to be same in both the cases. 

\vspace{.1in}
{\bf Region without Time-Reversal Symmetry, $(\beta=2)$:}
\vspace{.1in}

As $A_{nn}=\sum_{s=1}^2\sum_{k=1}^{2N} M_{nk;s}^2=
\sum_{r=0,N}\sum_{s=1}^2\sum_{k=1}^{N} M_{n k+r;s}^2$, we get 

\begin{eqnarray}
{\overline{\delta A_{nn} \delta A_{nn}^* }}  
&=& 4\sum_{r_1,r_2}\sum_{s_1,s_2}\sum_{k,l=1}^{N}
M_{n,k+r_1;s_1} M_{n,l+r_2;s_2} 
{\overline {(\delta M_{n,k+r_1;s_1} \delta M_{n,l+r_2;s_2})}}\\
&=& 8\delta Y \sum_{r_1,r_2}\sum_{s_1,s_2}\sum_{k,l=1}^{N}
 M_{n,k+r_1;s_1} M_{n,l+r_2;s_2} \delta_{k+r_1,l+r_2}\delta_{s_1,s_2}  \\
&=& 8\delta Y \sum_{r}\sum_{s}\sum_{k=1}^{N} M_{n,k+r;s}^2 \\
&=& 8  A_{nn} \delta Y.
\end{eqnarray}

Here eq.(D16) is obtained from eq.(D5) by using eq.(24) (or alternatively eq.(D7)).  	

\vspace{.1in}
{\bf Region with Time-Reversal Symmetry}
\vspace{.1in}

As $M_{nk;s}=(-1)^{s-1} M_{|n+N|,|k+N|;s}$ under time-reversal symmetry, 
$A_{nn}$  can be rewritten as 
$A_{nn}=\sum_{r=0,N}\sum_{s=1}^2\sum_{k=1}^{N} M_{n+r, k;s}^2$. This gives

\begin{eqnarray}
{\overline{\delta A_{nn} \delta A_{nn}^* }}  
&=& 4\sum_{r_1,r_2}\sum_{s_1,s_2}\sum_{k,l=1}^{N}
M_{n+r_1,k;s_1} M_{n+r_2,l;s_2} 
{\overline {(\delta M_{n+r_1,k;s_1} \delta M_{n+r_2,l;s_2})}}\\
&=& 4\delta Y \sum_{r_1,r_2}\sum_{s_1,s_2}\sum_{k,l=1}^{N}
 M_{n+r_1,k;s_1} M_{n+r_2,l;s_2} \left[\delta_{r_1,r_2}\delta_{kl}\delta_{s_1,s_2} 
+(-1)^{s_2-1} \delta_{r_1,r_2+N}\delta_{k,l+N}\delta_{s_1,s_2} \right ] \\
&=& 4\delta Y \sum_{r}\sum_{s}\sum_{k=1}^{N}
\left[ M_{n+r,k;s}^2 + (-1)^{s-1} M_{n+r+N,k+N;s} M_{n+r,k;s}\right] \\
&=& 8 \delta Y \sum_{r}\sum_{s}\sum_{k=1}^{N} M_{n+r,k;s}^2 \\
&=& 8  A_{nn} \delta Y
\end{eqnarray}
here eq.(D22) is obtained from eq.(D21) by using the relation 
$M_{n+r+N,k+N;s}={(-1)}^{s-1} M_{n+r,k;s}$.
 
	The eqs.(D9),*(D14) and (D23) can be combined together to give 
eq.(25).

\section{Proof of equation (10)}

	The matrix $A$ is Hermitian in nature. Let the inverse pair 
eigenvalues of $A$, at value $Y$ of the complexity parameter, 
be given by $x_n$ and $x_n^{-1}$, $n=1\rightarrow N$ at parametric values $Y$. 
A small change $\delta Y$ in parameter $Y$ changes $A$ and its eigenvalues. By 
considering matrix $A+\delta A$ in the diagonal representation of matrix $A$,  
a small change $\delta x_n$ in the eigenvalues can be given as 
\begin{eqnarray}
\delta x_n = \delta A_{nn} +\sum_{m\not=n} {|\delta A_{mn}|^2 \over x_n-x_m}+
o((\delta A_{mn})^3)
\end{eqnarray}
where $A_{mn}=x_n \delta_{mn}$ at value $Y$ of complexity parameter. 
This further gives, 
\begin{eqnarray}
\overline{\delta x_n} &=& \overline {\delta A_{nn}} +
\sum_{m=1,m\not=n}^{2N} {\overline{|\delta A_{mn}|^2} \over x_n-x_m} \\
&=& \left[ -2\gamma A_{nn} + 2 \sum_{m=1,m\not=n}^{2N} 
(\beta+(4-\beta)\delta_{m,n+N}) {A_{nn}+A_{mm} \over x_n-x_m}\right] \delta Y \\
&=& \left[ -2\gamma x_n + 2 \sum_{m=1,m\not=n}^{2N} 
(\beta +(4-\beta)\delta_{m,n+N}){x_n+x_m \over x_n-x_m}\right] \delta Y \\
&=& F_n(x) \delta Y
\end{eqnarray}
Here eq.(E3) in the above has been obtained from eq.(E2) by using eq.(25). 
Similarly, upto first order of $\delta Y$,   

\begin{eqnarray}
\overline{\delta x_n \delta x_m } =  
\overline{\delta A_{nn} \delta A_{mm}} =
8 A_{nn} \delta_{nm} \delta Y = 8  x_n \delta_{nm} \delta Y 
\end{eqnarray}

The eq.(E5) can now be used to obtain first moment of eigenvalues 
$\lambda_n={(x_n-1)^2\over 4x_n} $ of the matrix $B$. 

\begin{eqnarray}
\overline{\delta \lambda_n} =
{(x_n^2-1)\over 4 x_n^2}\overline{ \delta x_n}
={\sqrt{\lambda_n(1+\lambda_n)}\over x_n} F_n(\lambda) \delta Y
\end{eqnarray}

where 

\begin{eqnarray}
F_n &=& \left[ -2\gamma x_n +2 \beta  
\sum_{m=1,m\not=n}^{N} {(x_n+x_m)(x_n-x_m^{-1})+(x_n-x_m)(x_n+x_m^{-1})
 \over (x_n-x_m) (x_n-x_m^{-1})}
+ 8 {x_n+x_n^{-1} \over x_n-x_n^{-1}}\right]  \nonumber  \\
&=& {-1\over \sqrt{\lambda_n(1+\lambda_n)}}
[2\gamma x_n \sqrt{\lambda_n (1+\lambda_n)} 
- 4(2\lambda_n+1)
-4 \beta \sum_{m\not=n} {\lambda_n (1+\lambda_n)\over
\lambda_n-\lambda_m}]  \nonumber \\
&=& -4 {x_n \over \sqrt{\lambda_n (1+\lambda_n)} } E_n 
\end{eqnarray}
where $E_n$ is given by eq.(28).
The average of the second moment $(\delta \lambda_n)^2$ 
can similarly be calculated, 
\begin{eqnarray}
\overline{\delta \lambda_m \delta \lambda_n} =
{(x_m^2-1)\over 4 x_m^2}{(x_n^2-1)\over 4 x_n^2}
\overline{ \delta x_n \delta x_m}
=8 {\lambda_n(1+\lambda_n)\over x_n} \delta_{nm} \delta Y
\end{eqnarray}

\end{appendix}

%\end{multicols}

\begin{references}
\bibitem{vk1}
Note, although the diffusion coefficient in eq.(29) is position dependent,  
however the Ito-Strantanovich ambiguity of Brownian motion does not arise here. This 
is because eq.(27) explicitly relates the change in the variables $\lambda_n$ to 
their values prior to the change.
\end{references}
\end{document}